\documentclass[10pt, conference, compsocconf]{IEEEtran}

\usepackage{cite}

\ifCLASSINFOpdf
  \usepackage[pdftex]{graphicx}
\else
\fi

\usepackage[cmex10]{amsmath}
\begin{document}

\title{Direct Sequence Spread Spectrum Steganographic Scheme for IEEE 802.15.4}

\author{\IEEEauthorblockN{El\.zbieta Zieli\'nska and Krzysztof Szczypiorski}
\IEEEauthorblockA{Institute of Telecommunications\\
Warsaw University of Technology\\
Warsaw, Poland\\
\{E.Zielinska, K.Szczypiorski\}@tele.pw.edu.pl}}
\maketitle

\begin{abstract}
This work addresses the issues related to network steganography in IEEE 802.15.4 Wireless Personal Area Networks (WPAN). The proposed communication scheme employs illicit Direct Sequence Spread Spectrum code sequences for the transmission of steganographic data. The presented approach is a compromise between minimising the probability of covert channel disclosure and providing robustness against random errors and a high steganographic data rate. The conducted analyses show that it is possible to create a covert channel with a data rate comparable to the raw data rate of IEEE 802.15.4 without much impact on the perceived receiver sensitivity, the Chip Error Rate and the Bit Error Rate.
\end{abstract}


\IEEEpeerreviewmaketitle

\section{Introduction}

The popularisation of the IEEE 802.15.4 standard is propelled by the predicted implementation of the Internet of Things \cite{gershenfeld2004d} concept. The incorporation of Wireless Personal Area Networks (WPAN) into the IP-domain and the proliferation of IP-based sensor networks will trigger the same threats as presently observed in the Internet. Among these risks is steganography, named one of the more more significant security issues in present day networks \cite{intragency2006federal}.  

Steganography is a technique of conducting covert communication by means of embedding secret messages into some form of carrier \cite{fridrich2009steganography}. The carrier, in network steganography, may be defined in a variety of ways – the most common being the Protocol Data Unit (PDU) of a certain protocol but, generally, it may be any intrinsic property of a protocol or a number of protocols \cite{zander2007covert}. It is possible to utilise the time relations of the PDUs, their contents, losses, damages or ordering to convey the additional information. The main aim of steganography is to keep communication concealed so that unaware onlookers will not notice any aberration in comparison to standard protocol functioning.

\begin{figure}[!t]
\centering
\includegraphics[width=3.0in]{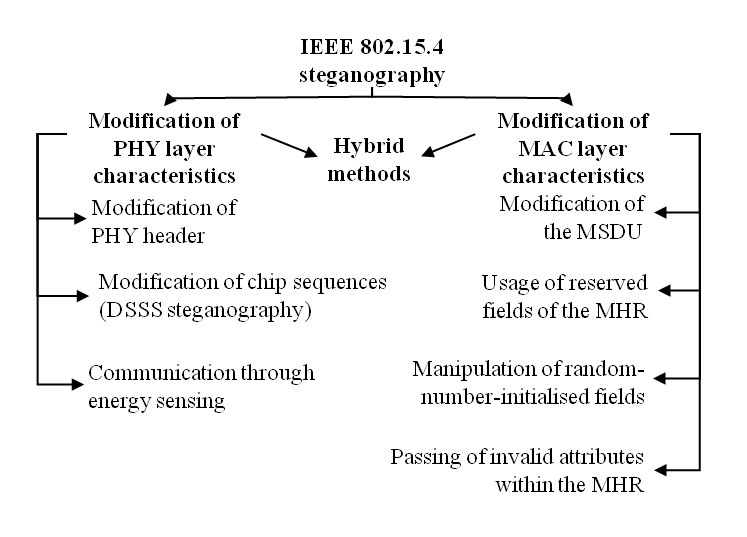}
\caption{Overview of steganographic techniques feasible in IEEE 802.15.4 networks.}
\label{fig:classification}
\end{figure}

Fig. \ref{fig:classification} is an illustration of various steganographic techniques which had been proposed for IEEE 802.15.4. The author of this work proposes a classification based on the International Organization for Standardization Open System Interconnection Reference Model (ISO OSI RM) layer, corresponding to the protocol, whose mechanisms are exploited for covert communication.
The physical layer steganography methods proposed for IEEE 802.15.4 represent three distinct approaches to the creation of covert channels. The first trend involves communication with the aid of Physical Layer (PHY) header fields. This low throughput method, proposed in \cite{martins2010attacks}, stipulates the usage of the PHY Service Data Unit (PSDU) length field in the PHY header to hide information.

The second group of methods, discussed in \cite{mehta2008steganography} and \cite{kho2007steganography}, involves the usage of Direct Sequence Spread Spectrum (DSSS) sequences to convey the message. The main principle of this approach is to tamper the standard provided DSSS sequences, and to transmit information over these induced "errors". The usage of this technique provides a very capacious covert channel, but it comes at a cost of deteriorating the efficiency of the underlying genuine IEEE 802.15.4 communication.

Lastly, there had been suggestions \cite{chebrolu2009esense} that it is possible to exploit the energy sensing mechanism to implement steganographic communication between devices within each other's energy sensing range. This can be possibly exploited by devices complying with different standards. The data set is a special alphabet consisting of different length packets. Here the information carrier is the energy burst duration. The main idea behind this scheme is the method for differentiation between regular packets and the steganographically modulated packets. This differentiation is possible thanks to the fact that the distribution of the length of regular packets is usually n-polar, where n is a small integer, and thus allows for the addition of a broad spectrum of different-length packets conveying additional meaning.

MAC layer allows for more versatility than the PHY in terms of the creation possibilities of hidden communication channels. The authors of \cite{martins2010 attacks} provide an overview of these methods, majority of which utilise the reserved fields of frames, or rely on the deliberate choice of values in otherwise random-initialised fields.
The first group of steganographic techniques utilises the reserved fields within the MAC header (MHR) to transmit up to 5 bits per packet with the aid of the reserved bits within the Frame Control field. The second group of method relies on the initialisation of otherwise randomly initialised fields with parts of, or whole, stego-objects. For example, this action can be performed on the Sequence Number field of the MAC header. Both of these methods are characterised by low steganographic bandwidth.

The rest of this work is organised as follows. Sections II and III provide an overview of steganographic methods for IEEE 802.15.4 and a detailed description of the DSSS steganography concept.
Section IV provides an embedding scheme in tampered DSSS sequences. Section V demonstrates the influence of covert channels on the performance metrics of underlying legit communication. Finally, section VI concludes the work.

\section{State of the Art}
Physical layer steganography, especially methods involving the modification of wideband transmission schemes, has proved to be suitable for the creation of high-bandwidth covert channels. Szczypiorski and Mazurczyk in \cite{szczypiorski2010hiding} have proposed a method for hiding data in Orthogonal Frequency Division Multiplexing (OFDM) symbols of IEEE 802.11 a/g/n networks, which provides a steganographic capacity of 1.12 Mb/s, in a channel with a raw data rate of 54 Mb/s. This method bases on the padding of OFDM data symbols.

DSSS codes' steganography is a promising method for covert communication in IEEE 802.15.4 networks, due to its capability to provide high-throughput channels. By far, the employment of this scheme permits for the creation of the most capacious of the hidden channels available for IEEE 802.15.4. As stated by Kho in \cite{kho2007steganography} and Mehta et al in \cite{mehta2008steganography}, it is feasible to exercise mechanisms characteristic to the IEEE 802.15.4 physical layer functioning in order to transmit hidden messages. According to these studies it is possible to create a steganographic channel with a bandwidth exceeding by over 6 times the raw data rate of IEEE 802.15.4 \cite{kho2007steganography}. 

The proposed solutions are based on the fact that every sequence consisting of four bits is mapped onto a sequence 32 chips-long, prior to its transmission. This procedure is referred to as symbol-to-chip mapping, where a symbol is the sequence of four bits. A chip is the most elementary unit of transmission – a modulation symbol, denoted with a 0 or a 1 and represented by a corresponding bipolar code of -1 or +1. These 32-chip sequences are predefined and constant for all IEEE 802.15.4 compliant networks. Since the transmission space is limited to 16 codes, it can be enlarged with little negative influence on network functioning. Addition of surplus sequences provides means of communication which can be utilised for steganographic message exchange.

Both works - \cite{mehta2008steganography} and \cite{kho2007steganography}, exploit an expanded alphabet of the DSSS sequences used to represent data symbols. The increase in the size of the steganographic alphabet results in the deterioration of receiver sensitivity.

\section{Principles of DSSS steganography in IEEE 802.15.4}
In IEEE 802.15.4, the modulation technique is 16-ary quasi-orthogonal, which means that the aforementioned 16-chip sequences are nearly orthogonal. The selection of these pseudo-random noise (PN) codes was performed with the aim of maximising the Hamming distance \cite{mackay2003information} between any two of the sequences. Upon erred reception, correction is performed by means of simple matching with one of the base sequences with a minimum Hamming distance to the received sequence.

An enlargement of the code space permits a steganographic node to communicate "illicit" signals, which can carry secret data. At the same time it does not deprive ordinary nodes of the ability to receive the original, underlying message. The presence of steganographic exchange can lead to an increase in the experienced Signal to Noise Ratio (SNR) which, in turn, causes an increase in the Bit Error Rate (BER). The stego-sequences' set may be interpreted as selectively negating some of the base sequences' chips, and thus inducing artificial "errors" carrying secret data. A base sequence, with the selectively negating stego-mask applied, may be treated as an additional code supplemented to the base, 16-ary set of DSSS codes.

\section{The proposed method}
The existing DSSS steganography methods for IEEE 802.15.4 are characterised by a strong correlation between the amount of data that can be embedded in the carrier and the observed degradation of transmission quality metrics. This is reflected by the relation between the IEEE 802.15.4 standard compliant set of sequences and the steganographic code set in terms of the Hamming distance values. The two possible DSSS steganography approaches involve, either maximising the Hamming distances between these sets, or, contrary, attempt to keep the stego-sequences close to their counterparts from the base set. In both cases, it is necessary to maximise distances between sequences originating from a common set.

\begin{figure}[!t]
\centering
\includegraphics[width=2.5in]{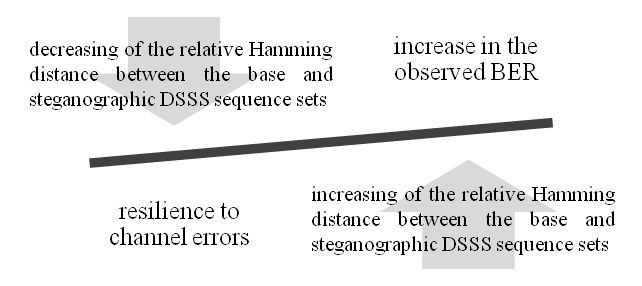}
\caption{The effect of the choice of steganographic DSSS sequence set on network performance.}
\label{fig:balance}
\end{figure}

The maximisation of the value of the minimum Hamming distance will allow for relatively errorless decoding of the received signals in poor channel conditions. An orthogonal approach is to intentionally provide steganographic sequences close to the default codes, which will allow for correct decoding, but the covert channel will be susceptible to channel errors. The general principle is illustrated in Fig. \ref{fig:balance}.
This work provides an algorithm for the embedding of additional content into IEEE 802.15.4 data symbols which ensures a high steganographic data rate while maintaining good performance characteristics. This is achieved with the knowledge of the properties of the base DSSS code set for IEEE 802.15.4 \cite{ieee2006standard}. These sequences are 32-chip-long, which gives a possible space of $2^{32}$ of allowed combinations. But this number is strongly limited. The sequences were chosen with the aim of maintaining maximum Hamming distances between the codes to provide good resistance to random errors. Calculations show that the minimum Hamming distance between a pair of codes, $d_{min}$ is 12, the average $\bar{d}$ is 17.1 while the maximum, $d_{max}$ is 20. It would be optimal to enlarge the set of existing codes in such a manner that the lower of these values remains unchanged. This would grant higher resistance to random errors and reduce the negative impact on carrier communication.

\begin{figure}[!t]
\centering
\includegraphics[width=3.0in]{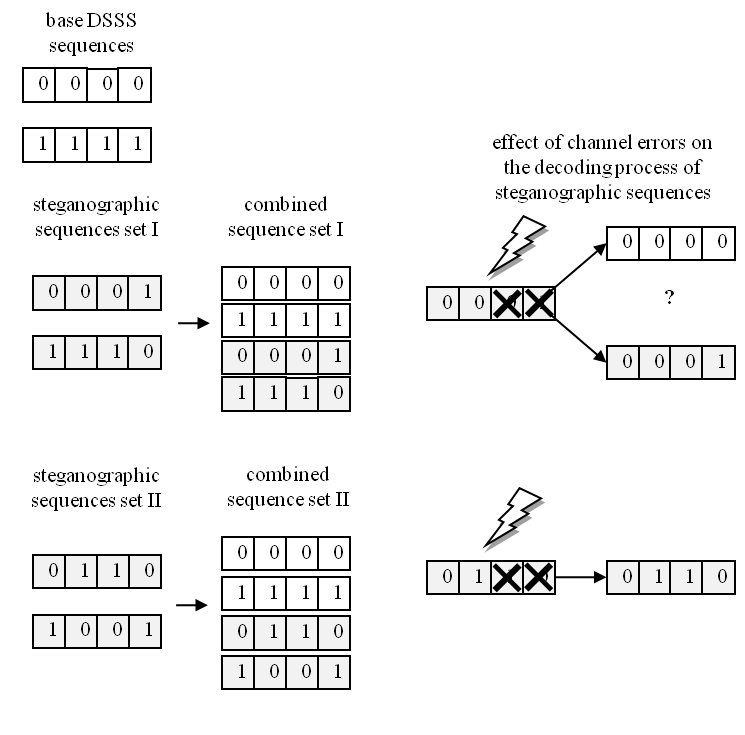}
\caption{A schematic example of the effect of choice of steganographic DSSS code selection on covert channel performance.}
\label{fig:mapping}
\end{figure}

Even with the application of the aforementioned rules governing the selection of steganographic DSSS sequences, the set of available codes is still very large. 
Fig. \ref{fig:mapping} provides a simplified example of such code selection illustrating the discussed relationship between Hamming distances and their influence on error resilience and the perceived BER. The first code set, whose Hamming distance to the base sequences is small, would camouflage efficiently the covert communication. On the other hand, it would also prove very susceptible to any errors naturally occurring in the medium. In such case, a steganographic sequence would be, with high probability, incorrectly mapped onto a symbol.

The second set, whose distance to the base sequences is by one larger than the prior's, is more resilient to random errors. This immunity comes at a cost of providing poorer camouflage of the steganographic information exchange. An onlooker, observing ongoing steganographic communication, would notice that the DSSS sequences in use do not match any of the default codes and would therefore infer that there is a suspicious increase in the Bit Error Rate.
From the point of view of providing covert system reliability, it seems natural to follow the Hamming distance maximisation approach, and this is the method of preference in this work. This choice will have effect on the algorithm for the generation of the steganographic sequences.

Let $d_{min}$ denote the minimum Hamming distance of the DSSS code set. Correct matching of a received sequence: $O$ with the sequence $s$ from the DSSS code set is always possible if the Hamming distance between the two: $d(O,s)<\lfloor\frac{d_{min}-1}{2}\rfloor$. In terms of IEEE 802.15.4, this signifies that in the absence of channel errors, it is possible to flip the values of up to 5 chips in any sequence without any increase in the Bit Error Rate (BER).

The altered chips in a sequence can convey steganographic payload with the utilisation of their position as the information carrier. This means that a steganographic receiver would map a received DSSS code onto the nearest sequence from the default set and thus obtain knowledge of the positions of the steganographically altered chips. Assuming that each layout of $i$ erroneous chips may be used as a unique steganographic sequence than the size of the covert communication alphabet may be defined as $N_{steg}$.
\begin{equation}
	N_{steg} = \sum\limits_{i=1}^{\lfloor\frac{d_{min}-1}{2}\rfloor} {32 \choose i}
\label{eq:stegset}
\end{equation}
This code space can be narrowed down if higher resistance of steganographic sequences to random errors is needed. For this purpose lets assume that the number of altered chips in every sequence is equal to $\lfloor\frac{d_{min}-1}{2}\rfloor$.
The proposed embedding procedure permits for the insertion of any of the steganographic symbols into any $s_i$ DSSS sequence belonging to the standard-compliant set $S$, $s_{i} \in S$. In order to decrease the probability of disclosure of the covert communication through autocorrelation of the observed signal it is necessary to employ a shared key to scramble the illicit code prior to the transmission. The scheme for the distribution of the keys used for scrambling between steganographic nodes is not discussed in this work. Let $e$ be the symbol that is to be transmitted covertly and $e_r$ its randomised version. The scrambling function is a pseudo-random permutation executed by a Linear Feedback Shift Register (LFSR) with the steganographic key - $k$ used as a seed. The pseudo-random permutation is a linear operation which can be inverted at the decoder to obtain the steganographic symbol. The employment of key rotation further increases security of the stego-scheme.
\begin{eqnarray}
e_r = Permute(e, k) \\
e = Inverse Permutation(e_r , k)
\label{eq:scrambling}
\end{eqnarray}
\section{Performance of IEEE 802.15.4 under stego-load}
The degradation of the perceived receiver sensitivity and the increase in the observed Bit Error Rate for IEEE 802.15.4 in the presence of steganographic communication is bound with the number of chip errors induced by the embedding process. Assuming that $n_{steg} ^{'}$ chips are altered in every 32-chip sequence, than the size of the steganographic code set: $N_{steg} ^{'} = {32 \choose \lfloor\frac{d_{min}-1}{2}\rfloor}$, which equals $201376$ under the considered constraints. If every sequence would map to one steganographic symbol, then every 32-chip sequence could carry an average of 17.62 steganographic bits. The cost of such an approach would be the lack of scrambling and poor error resilience of stego-symbols.
For the purpose of calculations, lets assume that every stego-sequence carries 4 data bits, which signifies that the steganographic data rate $R_{steg}$ is equal to the raw data rate of IEEE 802.15.4, $R = 250 kb/s$ \cite{ieee2006standard}. The rest of the code space is used for spreading purposes, therefore it may be considered that the probability that a certain chip of the carrier DSSS code is altered is constant, regardless of the chip position. Let $\Delta \bar{d}$ be the shift in the average Hamming distance, $\bar{d}$, triggered by the introduction of the chip alterations. 
\begin{equation}
	\Delta \bar{d} \approx \frac{\lfloor\frac{d_{min}-1}{2}\rfloor \cdot \bar{d}}{32} 
\label{eq:delta}
\end{equation}
Eq. \ref{eq:delta} reflects the reduction in relative distance between two randomly-selected codes from the IEEE 802.15.4 template DSSS codes' set as seen by a non-steganographic receiver.

\begin{figure}[!t]
\centering
\includegraphics[width=3.5in]{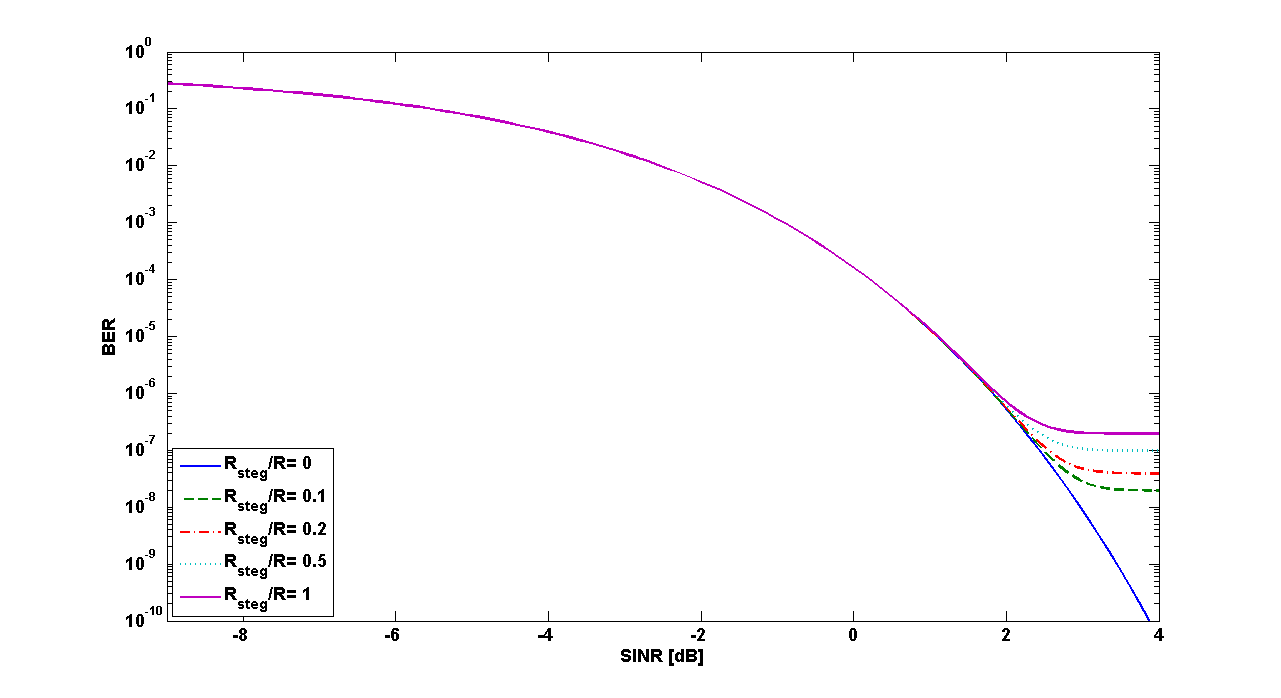}
\caption{BER experienced by a receiving node as seen by a non-steganographic receiver for different embedding rates.}
\label{fig:berplot}
\end{figure}

The observed shift in the value of BER can be estimated with the knowledge of the average number of bit errors in the event of incorrect sequence mapping and the shift in the probability of code misinterpretation (a symbol is 4 bits long).
\begin{equation}
\Delta BER = \frac{EX(num\ bit\ err\ per\ sym\ err)}{num\ bit\ per\ sym}\\ \cdot \Delta P_{m}
\label{eq:ber}
\end{equation}
The expected number of bit errors upon sequence mismatching (one of the 15 incorrect codes is picked) depends on the probabilities that 1 (4 sequences out of 15 match this condition), 2 (6 matching), 3 (4 matching) or 4 (1 matching) bits have been erroneous. Therefore, for IEEE 802.15.4, $EX(num\ bit\ err\ per\ sym\ err)$ equals $\frac{32}{15}$. 
$\Delta P_m$ is the change in the probability that a sequence is decoded incorrectly as a different code from the template set. It may be estimated with the knowledge of coding gain for minimum distance decoders. Assuming that $P_B$ is the probability of a bit error in a system without coding, $n$ is the code length (32 for IEEE 802.15.4), $t=\lfloor\frac{d_{min}-1}{2}\rfloor$ is the maximum error correcting capability, then the probability that a bit is erroneous in the presence of coding may be defined as $P_C$.
\begin{equation}
P_C \approx \frac{1}{n} \sum_{i=t+1}^{n} i \cdot {n \choose i} \cdot P_{B}^{i} (1-P_{B})^{n-i}  
\label{eq:bit}
\end{equation}
\newpage
From \ref{eq:bit}, where $P_{C steg}$ is the coding gain in the presence of ongoing covert communication, it is possible to calculate $\Delta P_m$.
\begin{equation}
\Delta P_{m} = \frac {P_{C steg}}{P_C}
\label{eq:deltapm}
\end{equation}
The standard-defined BER for IEEE 802.15.4 is presented in Eq. \ref{eq:berieee}.
\begin{equation}
BER = \frac{8}{15} \cdot \frac{1}{16} \sum_{k=2}^{16} {-1}^k {16 \choose k} e^{( 20 \cdot SNR \cdot (\frac{1}{k} - 1) )}
\label{eq:berieee}
\end{equation}
Fig. \ref{fig:berplot} illustrates the influence of steganographic communication on the error performance of IEEE 802.15.4-compliant device. The embedding rate $R_{steg}/R$ corresponds to the number of data symbols that are chosen from the data stream for embedding purposes. The steganographic modification of one data symbol (one DSSS sequence) yields 4 bits of steganographic capacity, therefore, for $R_{steg}/R = 1$ the covert channel bandwidth is 250 kb/s.

It is notable that the impact of the stego-system is most visible in good noise conditions. Assuming that the typical BER values for wireless networks fall in the range of $10^{-1}$ up to $10^{-6}$ \cite{garg2007wireless}, it may be deemed that the proposed stego-system functions below the noise floors for IEEE 802.15.4.
The deterioration of receiver performance is estimated basing on the characteristics of pure IEEE 802.15.4 communication. The increase in BER for a specific steganographic data rate can be projected onto a shift in SINR value in a pure system, corresponding to the same jump in BER values. The change of SINR can be interpreted as deterioration of receiver sensitivity (this is how a network device will interpret the change in characteristics).

\begin{figure}[!t]
\centering
\includegraphics[width=3.5in]{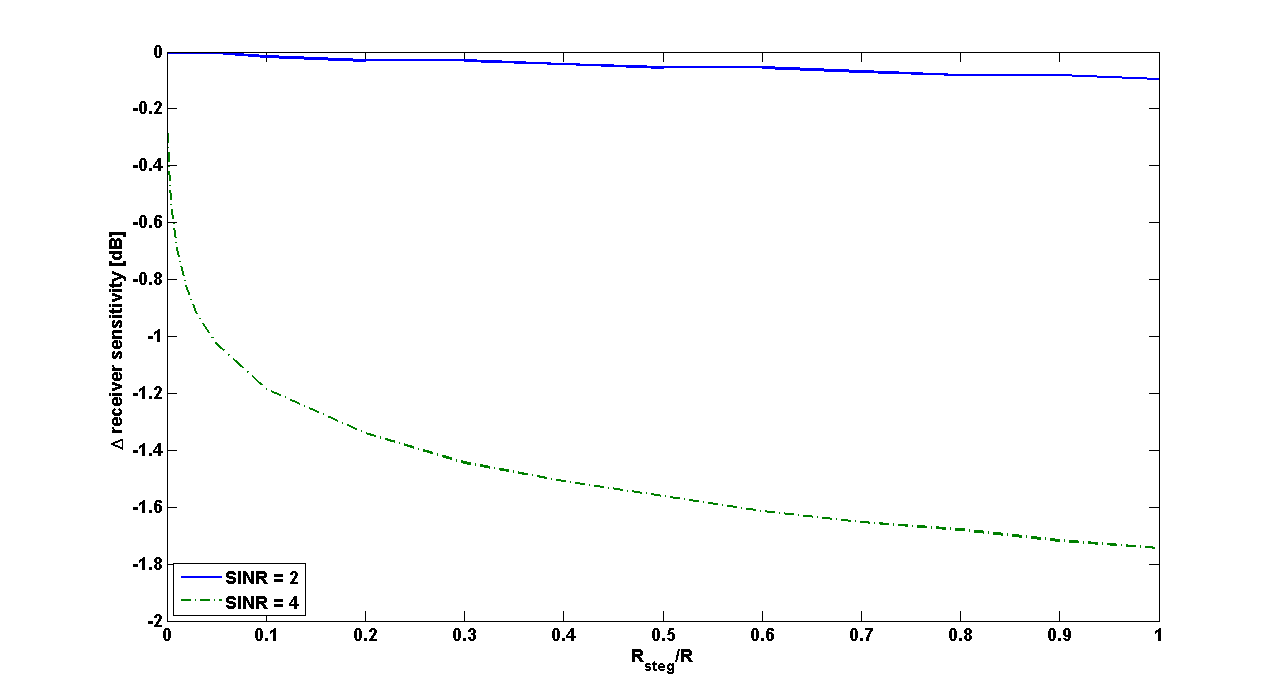}
\caption{The change in receiver sensitivity under the presence of different steganographic data rates and under different noise conditions.}
\label{fig:sensitivity}
\end{figure}

For the 2450 MHz PHY, a device must be capable of achieving a sensitivity of -85 dBm or better \cite{ieee2006standard}. Assuming that a receiver has a sensitivity of -100 dBm, the maximum change in sensitivity cannot exceed 15 dB. This phenomenon of apparent sensitivity decline in the presence of ongoing covert communication is illustrated in Fig. \ref{fig:sensitivity}.

The obtained results reveal that the proposed approach is more efficient than the one proposed in \cite{kho2007steganography}. For $SNR = 4dB$ and the embedding rate of 5 chips per symbol, the proposed scheme reduces the sensitivity of a receiver by 1.8dB, compared to over 8dB in \cite{kho2007steganography}. At the same time the covert bandwidth of the channel from \cite{kho2007steganography} is 312.5kb/s, compared to 250kb/s provided in the proposed solution.

\section{Conclusions}
The proposed DSSS steganographic coding scheme for IEEE 802.15.4 provides a method for the coding of steganographic data – this is achieved by means of introducing additional chip errors into the template DSSS codes. This algorithm is complimented by additional scrambling of stego-chips, which increases the resistance of the stego-system to detection by randomising the induced chip errors. The analysis of the effect of usage of such tampered codes on the decoding process revealed that it is possible to establish a 250 kb/s covert channel on top of the carrier communication with negligible impact on the Bit Error Rate, and a decrease in receiver sensitivity noticeable only in poor noise conditions (SNR exceeding 2 dB). The steganographic tampering of default DSSS code's chips is significantly mitigated by the correction capabilities of the receivers.

\newpage
\bibliographystyle{IEEEtran}
\bibliography{submission}

\end{document}